\documentclass[%
 reprint,
 amsmath,amssymb,
 aps,
]{revtex4-2}

\usepackage{graphicx}
\usepackage{dcolumn}
\usepackage{bm}
\usepackage{xcolor}
\begin{document}

\preprint{APS/123-QED}

\title{Dark Matter Deficient Galaxies as Probes of Dark Matter}

\author{Oem Trivedi $^{1}$, Abraham Loeb $^{2}$}
\affiliation{$^{1}$Department of Physics and Astronomy, Vanderbilt University, Nashville, TN, 37235, USA}
\affiliation{$^{2}$Astronomy Department, Harvard University, 60 Garden St., Cambridge, 02138, MA, USA}
\email{Email: oem.trivedi@vanderbilt.edu \\ Email : aloeb.cfa@harvard.edu }

\date{\today}

\begin{abstract}
Dark matter deficient galaxies provide a direct way to test whether baryons and dark matter can become sufficiently separated during galaxy evolution. We formulate a Dark Matter-Baryon Separability condition based on the relative incorporation efficiencies of the two components, requiring the final dark matter to baryon ratio to fall below an observationally defined threshold. Applied to high speed collisions, this condition constrains dark matter baryon momentum transfer cross section, interacting dark matter fraction and the efficiency of gravitational recapture. The same framework gives us bounds on the abundance and cooling time of dissipative dark matter, and on the integrated escape rate of ultralight fuzzy dark matter from shallow or tidally disturbed potentials. These results show how dark matter deficient galaxies can complement cosmological and laboratory probes by constraining late time dark matter interactions, dissipation and halo stability.
\end{abstract}

\maketitle

Dark matter is the dominant nonrelativistic matter component of the Universe and is essential to the standard picture of structure formation. Its cosmological abundance and gravitational effects are well constrained, but its microscopic nature remains unknown\cite{dm11rubin1970rotation}. Proposed candidates for it include weakly interacting massive particles, axions, sterile neutrinos, primordial compact objects, ultralight scalar fields, warm dark matter and particles belonging to self interacting or dissipative dark sectors \cite{dm1Cirelli:2024ssz,dm2Arbey:2021gdg,dm3Balazs:2024uyj,dm4Eberhardt:2025caq,dm5Bozorgnia:2024pwk,dm6Misiaszek:2023sxe,dm7OHare:2024nmr,dm8Adhikari:2022sbh,dm9Miller:2025yyx,dm10Trivedi:2025vry,hdmTrivedi:2025sbe}. There are certain conditions any such model should abide by, for example reproduce the observed dark matter abundance, behave approximately as a pressureless component during structure formation, have a sufficiently small sound speed and free streaming scale and remain stable over cosmological times. It should also satisfy existing limits on its interactions with baryons, radiation and other dark matter particles. \\

It is hence quite interesting to note that a small number of galaxies appear to contain much less dark matter than expected. The first widely discussed example was NGC 1052 DF2, whose low internal velocity dispersion was found to be consistent with the gravitational field produced by its stellar population alone \cite{df21van2018galaxy,df22danieli2019still,df23wasserman2018deficit}. NGC 1052 DF4 was later identified as a similar system in the same environment \cite{df41van2019second,df42montes2020galaxy,df43li2024rotation} and more recently, FCC 224 in the outskirts of the Fornax Cluster was found to share several properties with DF2 and DF4 \cite{fcc1romanowsky2024candidate,fcc2tang2025unexplained,fcc3buzzo2025new}. Yet another galaxy, NGC 1052 DF9, was reported to have a stellar velocity dispersion close to that expected from its stellar mass and well below the value predicted for a conventional dark matter halo\cite{df91keim2026third,df92gannon2023keck,df93keim2025kinematic,df94van2022trail,df95tang2026new}. These galaxies are generally diffuse, have low internal velocity dispersions and contain old, largely quiescent stellar populations. Some also host unusually luminous globular clusters or appear to be part of larger spatial and kinematic structures. Their most important common feature is that the dynamical mass inferred within the observed region is comparable to the baryonic mass already present in stars and gas. \\

This of course was the catalyst to some investigations and several mechanisms have been proposed to account for this deficiency. One possibility is that of strong tidal stripping, wherein a dwarf galaxy may begin with an ordinary dark matter halo but lose much of its extended and weakly bound dark matter during repeated close passages around a massive host. Since the stellar component is usually more compact, it can remain bound after a large fraction of the halo has been removed. A second possibility is tidal dwarf formation, wherein during a major galactic interaction, stars and gas from the progenitor disks can be drawn into tidal tails, while comparatively little of the dynamically hot halo dark matter enters the same debris. Bound condensations formed from this material may hence begin with a low dark matter content. A third possibility is that of a high speed bullet dwarf collision and in this case the gas shocks and slows, while the dark matter passes through more freely. New stellar systems may then form from the compressed gas with only a small fraction of the original dark matter. \\

These observations raise a broader question, which is that can the existence of dark matter deficient galaxies be translated into a general condition on dark matter models? A viable theory does not need to predict that such galaxies are common by any means. It must, however, allow at least some environments in which baryonic matter can be retained or assembled more efficiently than dark matter. This requires dark matter to remain sufficiently separable from the baryonic component during at least one plausible formation process. Previous studies have considered individual formation channels, tidal evolution and observational selection criteria. But these ingredients have not, to our knowledge, been combined into a single dimensionless parameter applicable across different dark matter models. In the rest of this Letter, we formulate these constraints. \\

Within an observational radius $R_{\rm obs}$, the baryonic mass can be written as
\begin{equation}
M_b(<R_{\rm obs})=M_\star(<R_{\rm obs})+M_{\rm gas}(<R_{\rm obs})
\end{equation}
For pressure supported systems, the dynamical mass near the three dimensional half light radius may then be estimated as
\begin{equation}
M_{\rm dyn}(<r_{1/2})\simeq\frac{3r_{1/2}\sigma_{\rm los}^2}{G}
\end{equation}
Using the projected half light radius $R_e$,
\begin{equation}
M_{\rm dyn}(<r_{1/2})\simeq\frac{4R_e\sigma_{\rm los}^2}{G}
\end{equation}
The dependence on $\sigma_{\rm los}^2$ makes the internal velocity dispersion the main dynamical quantity entering the inference. A low value of $\sigma_{\rm los}$ can make the measured mass consistent with the baryonic contribution even for an extended galaxy. The enclosed dark matter mass is defined by
\begin{equation}
M_\chi(<R_{\rm obs})=M_{\rm dyn}(<R_{\rm obs})-M_b(<R_{\rm obs})
\end{equation}
We now introduce the enclosed dark matter to baryon ratio
\begin{equation}
\mu_{\rm obs}\equiv\frac{M_\chi(<R_{\rm obs})}{M_b(<R_{\rm obs})}
\end{equation}
This means that a dark matter dominated galaxy has $\mu_{\rm obs}\gg1$, while a dark matter deficient galaxy would satisfy
\begin{equation}
\mu_{\rm obs}\leq\mu_{\rm crit}
\end{equation}
For definiteness, we adopt
\begin{equation}
\mu_{\rm crit}=1
\end{equation}
This choice so identifies systems in which the enclosed dark matter mass does not exceed the baryonic mass. The following derivation remains unchanged for any other observationally motivated value of $\mu_{\rm crit}$. Consider an initial progenitor with dark matter mass $M_\chi^{\rm i}$ and baryonic mass $M_b^{\rm i}$ and its initial mass ratio is
\begin{equation}
\mu_{\rm i}\equiv\frac{M_\chi^{\rm i}}{M_b^{\rm i}}
\end{equation}
For a formation channel $c$, define the fractions of the initial components incorporated into the final remnant as
\begin{equation}
\epsilon_{\chi c}\equiv\frac{M_{\chi c}^{\rm f}(<R_{\rm obs})}{M_\chi^{\rm i}}
\end{equation}
\begin{equation}
\epsilon_{bc}\equiv\frac{M_{bc}^{\rm f}(<R_{\rm obs})}{M_b^{\rm i}}
\end{equation}
For tidal stripping, these quantities are retained fractions and for tidal dwarf or bullet dwarf formation, they represent the fractions captured by the newly formed object. The final dark matter to baryon ratio is then given as
\begin{equation}
\mu_{{\rm f}c}\equiv\frac{M_{\chi c}^{\rm f}(<R_{\rm obs})}{M_{bc}^{\rm f}(<R_{\rm obs})}
\end{equation}
Substituting the incorporation fractions will end up giving us
\begin{equation}
\mu_{{\rm f}c}=\frac{\epsilon_{\chi c}M_\chi^{\rm i}}{\epsilon_{bc}M_b^{\rm i}}
\end{equation}
It is then straightforward to obtain
\begin{equation} \label{mufc}
\mu_{{\rm f}c}=\mu_{\rm i}\frac{\epsilon_{\chi c}}{\epsilon_{bc}}
\end{equation}
A deficient galaxy forms only when the final process incorporates baryons more efficiently than dark matter by an amount sufficient to overcome the initial dark matter dominance. The condition $\mu_{{\rm f}c}\leq\mu_{\rm crit}$ becomes
\begin{equation}
\frac{\epsilon_{\chi c}}{\epsilon_{bc}}\leq\frac{\mu_{\rm crit}}{\mu_{\rm i}}
\end{equation}
Thus, the relevant quantity is not the absolute amount of mass loss but the relative depletion or incorporation of the two components. If dark matter and baryons are affected in the same proportion, the final mass ratio remains unchanged. For an illustrative progenitor with $\mu_{\rm i}=100$ and $\mu_{\rm crit}=1$,
\begin{equation} \frac{\epsilon_{\chi c}}{\epsilon_{bc}}\leq10^{-2} \end{equation}
The dark matter fraction incorporated into the final object must therefore be at least two orders of magnitude smaller than the corresponding baryonic fraction. It is now useful to define $\delta\equiv\mu_{\rm crit}/\mu_{\rm i}$, so that the general requirement becomes simply $\epsilon_{\chi c}/\epsilon_{bc}\leq\delta$. This form allows the observed deficiency to be translated directly into constraints on the physical properties of dark matter. \\

We first consider dark matter scattering with baryons during a high velocity encounter. Let $f_{\rm int}$ denote the fraction of dark matter that participates in the interaction, while the remaining fraction behaves effectively collisionlessly. Then the momentum relaxation rate may be estimated as \cite{mom1dvorkin2014constraining,mom2xu2018probing}
\begin{equation} \Gamma_{\chi b}\simeq\rho_bv_{\rm rel}\frac{\sigma_{\chi b}^{\rm MT}(v_{\rm rel})}{m_\chi+m_b} \end{equation}
where $\sigma_{\chi b}^{\rm MT}$ is the momentum transfer cross section. Integrating this rate along the path through the baryonic material gives the drag depth
\begin{equation} D_{\chi b}\equiv\int\Gamma_{\chi b}dt\simeq\Sigma_b\frac{\sigma_{\chi b}^{\rm MT}(v_{\rm col})}{m_\chi+m_b} \end{equation}
Here $\Sigma_b$ is the baryonic column density and $v_{\rm col}$ is the characteristic collision velocity. The probability that an interacting dark matter particle undergoes sufficient momentum exchange can then be approximated by $1-e^{-D_{\chi b}}$. The collisionless fraction may still be incorporated into the final remnant through gravitational capture and for that, we denote this relative incorporation efficiency by $f_{\rm grav}$. \\

A simple phenomenological expression for the total dark matter incorporation relative to the baryons is then
\begin{equation} \frac{\epsilon_{\chi}}{\epsilon_b}\simeq f_{\rm int}\left(1-e^{-D_{\chi b}}\right)+\left(1-f_{\rm int}\right)f_{\rm grav} \end{equation}
The first term describes the interacting component that is slowed with the gas, while the second accounts for residual capture of the component that passes through the collision. The dark matter population is divided into an interacting fraction $f_{\rm int}$ and an effectively collisionless fraction $1-f_{\rm int}$. Since $D_{\chi b}$ is the integrated interaction depth, $e^{-D_{\chi b}}$ is the probability of undergoing no effective momentum exchange so $1-e^{-D_{\chi b}}$ gives the fraction slowed with the gas. The two mutually exclusive contributions are then weighted by their abundances and added, with $f_{\rm grav}$ representing the relative gravitational capture efficiency of the collisionless component. The dark matter deficient galaxy condition hence becomes
\begin{equation} f_{\rm int}\left(1-e^{-D_{\chi b}}\right)+\left(1-f_{\rm int}\right)f_{\rm grav}\leq\delta \end{equation}
This inequality is the main physical constraint obtained from the collision scenario and we can term it as the Dark Matter-Baryon Separability criterion. It places a joint limit on the interacting fraction, the dark matter baryon momentum transfer cross section and the efficiency of gravitational recapture. Now, this can be applied directly to the parameters of a microscopic dark matter model. \\

Consider now that the full dark matter component interacts, so $f_{\rm int}=1$, and then the condition reduces to $1-e^{-D_{\chi b}}\leq\delta$. This ends up giving us
\begin{equation} \frac{\sigma_{\chi b}^{\rm MT}(v_{\rm col})}{m_\chi+m_b}\leq\frac{-\ln\left(1-\delta\right)}{\Sigma_b} \end{equation}
For $\delta=10^{-2}$ and a realistic but illustrative value $\Sigma_b=0.02\,{\rm g\,cm^{-2}}$, we find
\begin{equation} \frac{\sigma_{\chi b}^{\rm MT}(v_{\rm col})}{m_\chi+m_b}\lesssim0.50\,{\rm cm^2\,g^{-1}} \end{equation}
This estimate should be regarded as conditional on the adopted collision geometry and baryonic column density. A more precise limit will require a reconstruction or simulation of the relevant encounter, including the impact parameter, gas structure and gravitational capture of the passing dark matter. An opposite limit gives a useful constraint on the interacting fraction, as if $D_{\chi b}\gg1$, then we can state that the interacting component is efficiently slowed. In that case, the inequality becomes
\begin{equation} f_{\rm int}\leq\frac{\delta-f_{\rm grav}}{1-f_{\rm grav}} \end{equation}
provided that $f_{\rm grav}<\delta$. If gravitational recapture is negligible, this gives $f_{\rm int}\lesssim\delta$ or $f_{\rm int}\lesssim1\%$ for the illustrative progenitor considered above. Even a strongly interacting dark component can therefore remain viable if it represents only a sufficiently small fraction of the total dark matter. Note that velocity independent dark matter baryon scattering is already tightly constrained by the cosmic microwave background and the matter power spectrum, since significant coupling around recombination would transfer baryonic pressure and photon diffusion effects to the dark matter. The present constraint is hence most useful for interactions that are enhanced at late times, become important only at galactic collision velocities, involve a subdominant component or possess inelastic or dissipative thresholds. Our illustrative strong drag result $f_{\rm int}\lesssim5\times10^{-3}$ is particularly notable because CMB sensitivity to a tightly baryon coupled component has also been found to weaken near the subpercent level, although the two constraints apply at very different epochs and velocities \cite{first1,first2}. We can write a general velocity dependent interaction as \cite{mom1dvorkin2014constraining,mom3maamari2021bounds}
\begin{equation} \sigma_{\chi b}^{\rm MT}(v)=\sigma_0\left(\frac{v}{v_0}\right)^n \end{equation}
For $f_{\rm int}=1$ and negligible gravitational recapture, the corresponding bound is
\begin{equation} \frac{\sigma_0}{m_\chi+m_b}\leq\frac{-\ln\left(1-\delta\right)}{\Sigma_b}\left(\frac{v_0}{v_{\rm col}}\right)^n \end{equation}
For $m_\chi\gg m_b$, the denominator may be approximated by $m_\chi$ and so, the positive values of $n$ are particularly interesting in this case because they allow the interaction to remain weak at recombination while becoming stronger during a high velocity galactic encounter. \\

\begin{figure*}[t]
\centering
\includegraphics[width=0.8\textwidth]{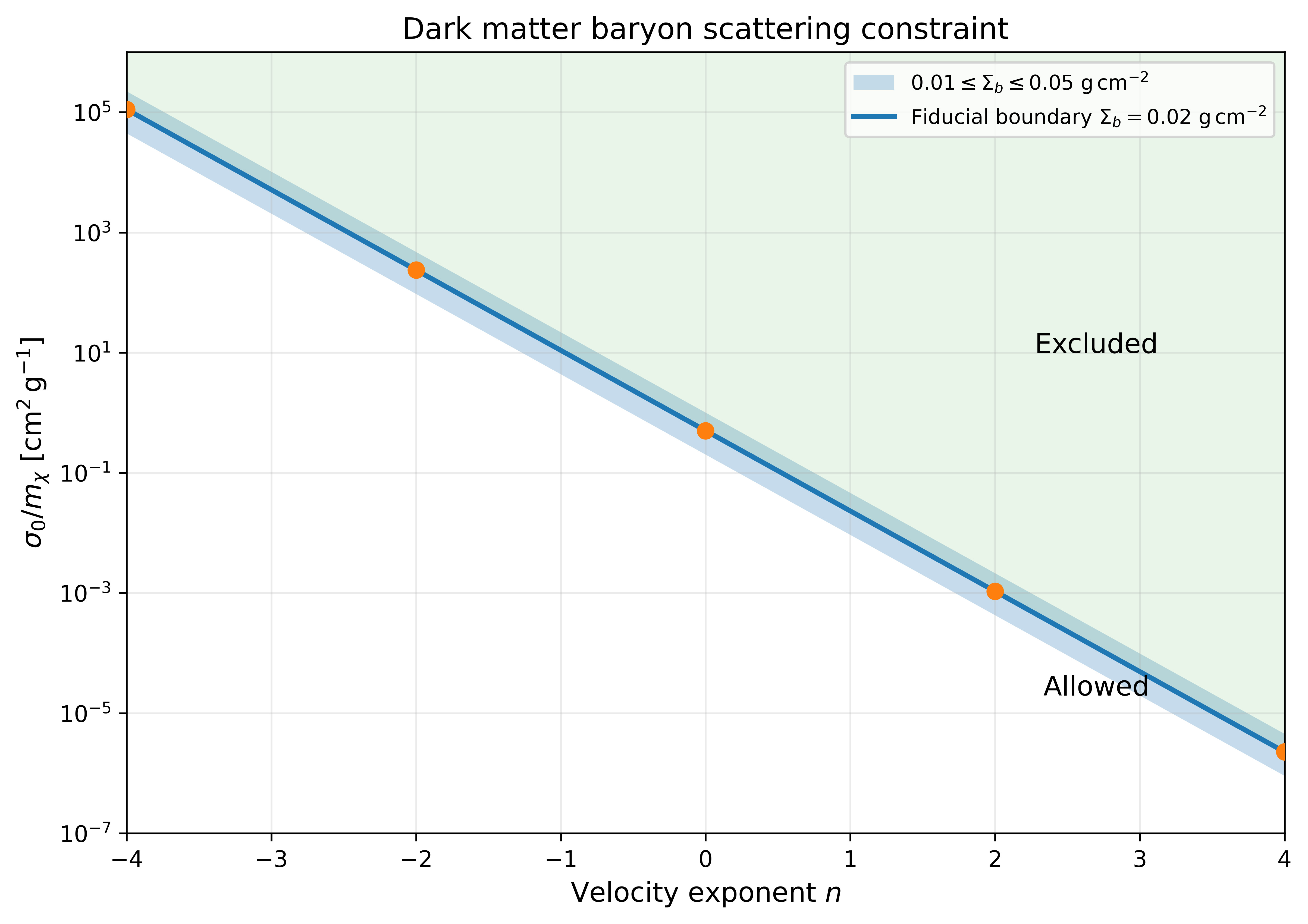}
\caption{The solid curve shows the upper bound on $\sigma_0/m_\chi$ as a function of the velocity exponent $n$, assuming $f_{\rm int}=1$, negligible gravitational recapture, $\delta=10^{-2}$, $v_0=30\,{\rm km\,s^{-1}}$, and $v_{\rm col}=651\,{\rm km\,s^{-1}}$.}
\label{sepplot}
\end{figure*}

To proceed further, we would like to consider an exclusion plot in the $\sigma_0/m_\chi$ versus $n$ plane and we have done just that in figure \ref{sepplot}.
The solid curve shows the upper bound on $\sigma_0/m_\chi$ as a function of the velocity exponent $n$, assuming $f_{\rm int}=1$, negligible gravitational recapture, $\delta=10^{-2}$, $v_0=30\,{\rm km\,s^{-1}}$ and $v_{\rm col}=651\,{\rm km\,s^{-1}}$. The shaded band reflects the variation obtained for $0.01\leq\Sigma_b\leq0.05\,{\rm g\,cm^{-2}}$, with the central curve corresponding to $\Sigma_b=0.02\,{\rm g\,cm^{-2}}$. Values above the boundary are excluded because they would couple dark matter too strongly to the baryonic remnant, while values below it satisfy the deficiency condition. The bound becomes rapidly stronger for positive $n$, reaching $\sigma_0/m_\chi\lesssim1.1\times10^{-3}$ and $2.3\times10^{-6}\,{\rm cm^2\,g^{-1}}$ for $n=2$ and $n=4$, respectively. \\

We can go further by retaining both the interacting fraction and gravitational recapture in a general condition. Solving
$f_{\rm int}(1-e^{-D_{\chi b}})+(1-f_{\rm int})f_{\rm grav}\leq\delta$
for the drag depth gives
\begin{equation}
D_{\chi b}\leq-\ln\left[1-\frac{\delta-(1-f_{\rm int})f_{\rm grav}}{f_{\rm int}}\right]
\end{equation}
which means
\begin{equation}
\frac{\sigma_{\chi b}^{\rm MT}(v_{\rm col})}{m_\chi+m_b}\leq-\frac{1}{\Sigma_b}\ln\left[1-\frac{\delta-(1-f_{\rm int})f_{\rm grav}}{f_{\rm int}}\right]
\end{equation}
This result is valid when $\delta>(1-f_{\rm int})f_{\rm grav}$ and if this condition is not satisfied, gravitational recapture alone incorporates too much dark matter and no value of the scattering cross section can produce the required deficiency. In the optically thin limit, this bound would then reduce to
\begin{equation}
f_{\rm int}\Sigma_b\frac{\sigma_{\chi b}^{\rm MT}}{m_\chi+m_b}+(1-f_{\rm int})f_{\rm grav}\lesssim\delta
\end{equation}
Thus the relevant observable combination is approximately $f_{\rm int}\sigma_{\chi b}^{\rm MT}/(m_\chi+m_b)$, rather than the cross section alone. Taking $\delta=10^{-2}$, $\Sigma_b=0.02\,{\rm g\,cm^{-2}}$ and $f_{\rm grav}=5\times10^{-3}$ gives upper limits of approximately $0.50$, $0.76$ and $2.83\,{\rm cm^2\,g^{-1}}$ for $f_{\rm int}=1$, $0.5$ and $0.1$, respectively. For $f_{\rm int}=1$, $m_\chi\gg m_b$, $v_0=30\,{\rm km\,s^{-1}}$ and $v_{\rm col}=651\,{\rm km\,s^{-1}}$, the velocity dependent bound gives $\sigma_0/m_\chi\lesssim0.50\,{\rm cm^2\,g^{-1}}$ for $n=0$, $\sigma_0/m_\chi\lesssim1.1\times10^{-3}\,{\rm cm^2\,g^{-1}}$ for $n=2$ and $\sigma_0/m_\chi\lesssim2.3\times10^{-6}\,{\rm cm^2\,g^{-1}}$ for $n=4$. A complementary limit follows in the strong drag regime $D_{\chi b}\gg1$, where
\begin{equation}
f_{\rm int}\leq\frac{\delta-f_{\rm grav}}{1-f_{\rm grav}}
\end{equation}
For the same illustrative values as above, this gives $f_{\rm int}\lesssim5.0\times10^{-3}$. This means that if the interaction is strong enough to lock dark matter to the gas, no more than about half a percent of the total dark matter may belong to that component. These numerical values remain conditional on the adopted column density and recapture efficiency, but they show how dark matter deficient galaxies can separately constrain the interaction strength and the interacting dark matter fraction. Note that unlike cosmological and cluster thermodynamic bounds, which constrain cumulative heat and momentum exchange, the present result limits the interaction through the requirement that dark matter not be incorporated into a particular shocked baryonic remnant \cite{second1,second2} \\

Related constraints can also apply to dissipative dark matter \cite{diss1foot2015dissipative,diss2essig2019constraining}. To demonstrate that, let $f_{\rm dd}$ denote the fraction of the progenitor dark matter that cools into a compact disk or another baryon following component. We define $\eta_{\rm dd}$ as the incorporation efficiency of this compact component relative to the baryons and $\eta_h$ as the corresponding efficiency for the remaining extended halo. The total relative incorporation is then given as
\begin{equation} \frac{\epsilon_\chi}{\epsilon_b}=f_{\rm dd}\eta_{\rm dd}+\left(1-f_{\rm dd}\right)\eta_h \end{equation}
Requiring this ratio to remain below $\delta$ gives us
\begin{equation} f_{\rm dd}\leq\frac{\delta-\eta_h}{\eta_{\rm dd}-\eta_h} \end{equation}
This result applies when $\eta_{\rm dd}>\eta_h$ and $\eta_h<\delta$. If the dissipative component follows the baryonic disk efficiently,$\eta_{\rm dd}\simeq1$, while the extended halo contributes little to the tidal remnant $\eta_h\ll\delta$ and then the constraint becomes
\begin{equation} f_{\rm dd}\lesssim\delta \end{equation}
For $\delta=10^{-2}$, no more than roughly one percent of the progenitor dark matter could occupy a component that follows the baryonic disk into the deficient remnant. This is a conditional but physically transparent constraint on dissipative dark matter models. A model in which all of the dark matter cools into a compact configuration would tend to carry dark matter into tidal debris or shocked gas and would therefore have difficulty producing the observed deficiency. The constraint is here much weaker when the dissipative sector is only subdominant and the dominant component remains extended and effectively collisionless. Moving forward, we can relate the dark disk fraction to the underlying cooling efficiency as well. Let $f_{\rm diss}$ denote the total fraction of dark matter belonging to the dissipative sector, then a simple estimate for the fraction that cools into the compact component over a time $t_{\rm age}$ is
\begin{equation}
f_{\rm dd}=f_{\rm diss}\left(1-e^{-t_{\rm age}/t_{{\rm cool}\chi}}\right)
\end{equation}
Substituting this into the general incorporation bound gives
\begin{equation}
f_{\rm diss}\left(1-e^{-t_{\rm age}/t_{{\rm cool}\chi}}\right)\leq\frac{\delta-\eta_h}{\eta_{\rm dd}-\eta_h}
\end{equation}
This constrains the dissipative abundance and cooling time simultaneously in a very interesting way. For $\delta=10^{-2}$, $\eta_{\rm dd}=1$ and an illustrative halo overlap $\eta_h=10^{-3}$, the right hand side is approximately $9.0\times10^{-3}$. In the rapid cooling limit, $t_{{\rm cool}\chi}\ll t_{\rm age}$, this gives us $f_{\rm diss}\lesssim0.9\%$. Conversely, when $f_{\rm diss}$ exceeds this value, the cooling time must satisfy
\begin{equation}
t_{{\rm cool}\chi}\geq\frac{t_{\rm age}}{-\ln\left[1-\frac{\delta-\eta_h}{f_{\rm diss}\left(\eta_{\rm dd}-\eta_h\right)}\right]}
\end{equation}
For $t_{\rm age}=10\,{\rm Gyr}$ and $f_{\rm diss}=0.1$, the illustrative values above require $t_{{\rm cool}\chi}\gtrsim106\,{\rm Gyr}$. A ten percent dissipative sector would hence need to cool very inefficiently in order to remain compatible with the formation of the deficient remnant. Since $t_{{\rm cool}\chi}$ can be calculated from the dark sector density, temperature and cooling function, this result provides a direct route from the observed deficiency to constraints on dissipative dark matter microphysics. It should be noted that Gaia stellar kinematics constrain the local surface density and scale height of a cooled dark disk, while the deficient galaxy condition instead constrains the cosmological dissipative fraction and cooling time through the requirement that such a component not follow baryons into the remnant \cite{third1,third2}. \\

Ultralight or fuzzy dark matter \cite{fdm1hu2000fuzzy,fdm2bar2019relaxation} provides a different application for this criterion as well. In this case, dark matter need not couple directly to the gas but its wave supported halo or solitonic core may lose mass through tidal escape and tunnelling from a shallow gravitational potential. We describe dark matter mass loss in this case through an effective escape rate
\begin{equation} \frac{dM_\chi}{dt}=-\Gamma_{\rm esc}M_\chi \end{equation}
The retained dark matter fraction after the relevant interaction is
\begin{equation} \epsilon_\chi=\exp\left[-\int\Gamma_{\rm esc}(t)dt\right] \end{equation}
Combining this result with $\epsilon_\chi/\epsilon_b\leq\delta$ gives the minimum integrated escape required to form a dark matter deficient remnant
\begin{equation} \int\Gamma_{\rm esc}(t)dt\geq\ln\left(\frac{\mu_{\rm i}}{\mu_{\rm crit}\epsilon_b}\right) \end{equation}
For $\mu_{\rm i}=100$, $\mu_{\rm crit}=1$ and a baryonic retention fraction $\epsilon_b=0.25$, this becomes
\begin{equation} \int\Gamma_{\rm esc}(t)dt\gtrsim\ln\left(400\right)\simeq5.99 \end{equation}
The fuzzy dark matter component must hence lose nearly six exponential e-folding factors of its initial mass during the interaction while leaving a bound stellar remnant. This should be accompanied by a pre-encounter survival condition given as
\begin{equation} \int_{\rm pre}\Gamma_{\rm esc}(t)dt\lesssim1 \end{equation}
The two conditions require the halo to remain stable before the encounter but become rapidly depleted during it. Since $\Gamma_{\rm esc}$ depends on the ultralight particle mass, soliton density, external tidal field and orbital history, numerical Schrodinger-Poisson evolution can translate these inequalities into a constrained range of the particle mass. The dark matter deficient galaxy then supplies a required minimum mass loss, while the survival of the progenitor supplies an upper limit on the earlier escape rate. Even without building full simulations as that is beyond the scope of the current paper, we can obtain further conditional constraints by separating the long pre-encounter evolution from the short period of strong tidal forcing. Approximating the escape rates in these two stages as $\Gamma_{\rm pre}$ and $\Gamma_{\rm enc}$, the survival and depletion requirements imply
\begin{equation}
\Gamma_{\rm pre}t_{\rm pre}\lesssim1\qquad\Gamma_{\rm enc}t_{\rm enc}\gtrsim\ln\left(\frac{\mu_{\rm i}}{\mu_{\rm crit}\epsilon_b}\right)
\end{equation}
It follows that
\begin{equation}
\frac{\Gamma_{\rm enc}}{\Gamma_{\rm pre}}\gtrsim\ln\left(\frac{\mu_{\rm i}}{\mu_{\rm crit}\epsilon_b}\right)\frac{t_{\rm pre}}{t_{\rm enc}}
\end{equation}
For $\mu_{\rm i}=100$, $\mu_{\rm crit}=1$, $\epsilon_b=0.25$, $t_{\rm pre}=8\,{\rm Gyr}$ and $t_{\rm enc}=0.5\,{\rm Gyr}$, the escape rate must increase by a factor of at least $\sim96$ during the encounter. The required deficiency will hence select models in which tidal escape is weak over most of the progenitor lifetime but becomes rapidly efficient during a close passage. A complementary condition can also follow from the stability of an FDM soliton. We note that numerical studies find runaway disruption once the ratio of the soliton central density to the mean host density inside the orbit falls below a value of order $4.5$. If we define $\mathcal R_\rho=\rho_c/\bar{\rho}_{\rm host}$, one therefore requires
\begin{equation}
\mathcal R_{\rho,{\rm pre}}\gtrsim4.5\qquad\mathcal R_{\rho,{\rm enc}}\lesssim4.5
\end{equation}
This provides a direct condition on the soliton density and the orbital environment, and shows that the progenitor must remain sufficiently dense during its earlier evolution but cross the disruption threshold during the encounter that produces the deficient remnant. We note here that a useful semianalytic estimate can relate the survival of a solitonic core to the particle mass \cite{sol1du2018tidal,sol2schive2014cosmic}
\begin{multline}
M_c\gtrsim5.82\times10^8\left[\mathcal R_{\min}(N)\right]^{1/4}m_{22}^{-3/2}\left(\frac{D}{\rm kpc}\right)^{-3/4} \\ \left(\frac{M_h}{10^{12}M_\odot}\right)^{1/4}M_\odot
\end{multline}
where $m_{22}=m_\phi/(10^{-22}\,{\rm eV})$, $D$ is the orbital distance and $\mathcal R_{\min}(N)$ is the minimum density ratio required to survive $N$ orbits. Applying this condition before the encounter and reversing it for disruption during the encounter gives us a two sided estimate
\begin{equation}
\left(\frac{A_{\rm pre}}{M_c}\right)^{2/3}\lesssim m_{22}\lesssim\left(\frac{A_{\rm enc}}{M_c}\right)^{2/3}
\end{equation}
where $A$ denotes the coefficient on the right hand side of the preceding relation without the factor $m_{22}^{-3/2}$. For illustrative purposes we take $M_c=10^8M_\odot$, $M_h=10^{12}M_\odot$, $D_{\rm pre}=100\,{\rm kpc}$, $D_{\rm enc}=20\,{\rm kpc}$, $\mathcal R_{\min}(10)=74$ and $\mathcal R_{\min}(1)=8.4$ gives
\begin{equation}
6.6\times10^{-23}\,{\rm eV}\lesssim m_\phi\lesssim1.0\times10^{-22}\,{\rm eV}
\end{equation}
Note that this interval is only illustrative and depends strongly on the core mass, host profile and orbital history. Nevertheless, it shows that the simultaneous requirements of early survival and encounter driven depletion can produce a finite particle mass window even before carrying out a dedicated Schrodinger Poisson simulation. A complete treatment must additionally of course verify that the stellar remnant remains bound while the fuzzy dark matter core is disrupted. Note that the illustrative disruption window obtained here lies below the lower masses generally favored by Lyman alpha forest and ultra faint dwarf heating constraints, so it should presently be interpreted as a conditional tidal requirement rather than an independently viable pure fuzzy dark matter interval \cite{fourth1,fourth2}. \\

The constraints above show the constraining power of the separability condition. The same observational requirement can be expressed as a bound on the dark matter baryon momentum transfer cross section, the fraction of strongly interacting dark matter, the fraction of a compact dissipative component or the integrated escape rate of an ultralight halo. The derived inequalities here provide the quantities that can be confronted directly with particle models. At present these limits remain conditional on the formation history of the observed galaxies and their practical strength will improve as simulations determine the relevant column densities, capture fractions, tidal histories and baryonic retention efficiencies. Unlike conventional bounds that are tied to a particular epoch or observable, the separability condition provides a common late time framework for translating the existence of these galaxies into constraints on several distinct classes of dark matter physics. Our main novelty here is that the same fundamental requirement can be mapped onto interaction strength, interacting fraction, cooling efficiency or halo escape, depending on the microscopic model being tested. This makes the framework readily extendable to new formation channels and dark sector models as improved simulations and observations become available. A larger observational sample will also make it possible to move beyond the existence of a single allowed event and constrain the predicted frequency of deficient systems. In this way, dark matter deficient galaxies can provide a late time and environmentally distinct complement to cosmological, laboratory and conventional halo constraints on dark matter physics. The main point we would like to drive home is that the possible (perhaps inevitable) sightings of more Dark Matter deficient galaxies is perhaps not just a series of astrophysical oddities but instead, could collectively constitute a new way to constrain dark matter physics itself. \\

\textbf{Acknowledgements}: OT was supported in part by the Vanderbilt Discovery Doctoral Fellowship. The work of AL is supported in part by the Black Hole Initiative, which is funded by GBMF and JTF.
\bibliography{apssamp}

\bibliographystyle{apsrev4-2}
\end{document}